\documentclass[prl,twocolumn,aps,superscriptaddress,preprintnumbers,letterpaper]{revtex4}
\usepackage{amsmath,amssymb}
\usepackage{epsfig}
\usepackage{graphicx}
\usepackage{amsmath}
\usepackage{amsfonts}
\usepackage{epstopdf}
\newcommand{\be}{\begin{equation}}
\newcommand{\ee}{\end{equation}}
\newcommand{\bear}{\begin{eqnarray}}
\newcommand{\eear}{\end{eqnarray}}
\newcommand{\ba}{\begin{array}}
\newcommand{\ea}{\end{array}}

\begin{document}

\title{Holographic Pomeron and Entropy}
\author{Alexander Stoffers and Ismail Zahed}
\affiliation{Department of Physics and Astronomy, Stony Brook University, Stony Brook, New York 11794-3800, USA}

\date{\today}

\begin{abstract}
In dipole-dipole scattering at large rapidity $\chi={\rm ln}(s/s_0)$, the induced instanton on the string worldsheet carries entropy ${\bf S}_k=2(\alpha_{{\bf P}k}-1)\chi$ with $\alpha_{{\bf P}k}-1$ the pomeron intercept for a dipole source of N-ality $k$. This stringy entropy is neither coherent nor thermal. We argue that it is released promptly over a time $t_R\approx  ({\bf b} / \chi)^3 / (4 \alpha') $ with $\alpha'/2$ the pomeron slope and ${\bf b}$ the impact parameter. This stringy entropy may explain the $3/2$ jump in the total charged multiplicities at about 10 participants reported over a wide range of collider energies by PHOBOS. We predict the charged multiplicities in $pp$, $pA$ and central $AA$ collisions at LHC.
\end{abstract}

\maketitle

\setcounter{footnote}{0}

\vskip0.2cm

{\bf 1.\,\,Introduction.}
The issue of how entropy is released in hadron-hadron and nucleus-nucleus collisions is a fundamental problem in the current heavy-ion program at collider energies. How coherence, which is a hallmark of a fundamental collision, turns to incoherence, which is at the origin of the concept of entropy, is a theoretical question of central importance. A possible understanding for the entropy  deposition was attempted at weak coupling through the concept of the color glass approach in classical but perturbative QCD \cite{Kharzeev:2001yq,Baier:2002bt,Fries:2008vp} and at strong coupling through the concept of black hole formation in holographic QCD  \cite{Shuryak:2005ia,Gubser:2008pc,Lin:2009pn,Wu:2011yd,Kiritsis:2011yn}.

The evidence of a strongly coupled plasma released at collider energies, with large and prompt entropy deposition and flow, suggest that a strong coupling approach is needed for the mechanism of entropy decomposition. In this way, the holographic approach with the release of a black hole falling along the holographic direction provides a plausible mechanism for entropy production. However, this mechanism is detached from our understanding of fundamental $pp$ collisions, which are after all the seeds at the origin of the entropy production. This note is an attempt to provide such an understanding.

$pp$ collisions at large rapidity are dominated by pomeron and reggeon exchange, \cite{GRIBOV,Donnachie:1992ny}.  At large $N_c$ and strong coupling, the pomeron exchange has a simple holographic realization as non-critical closed string exchange in the t-channel in $D=5$~\cite{Basar:2012jb}.  For a description of the pomeron as a closed string exchange in critical  $D=10$ dimensions  using the Virasoro-Shapiro string amplitude see~\cite{Brower:2006ea}.

At large rapidity $\chi$, this string
exchange is characterized by an effective Unruh temperature. This temperature is caused by the emergence of a longitudinal acceleration of the string caused by a global and longitudinal 'electric field' on the string world-sheet. This global electric field encodes twisted boundary conditions, and gives rise to a stringy instanton as the pomeron in dipole-dipole scattering at large rapidity. Below we suggest that the Unruh temperature causes the string to partially vibrate and thus carry entropy. 

In section 2, we revisit the arguments for the emergence of a stringy instanton presented in \cite{Basar:2012jb} for dipole-dipole scattering. In section 3, we derive the entropy associated to the stringy instanton  and tie it with the wee-dipole multiplicity characteristic of the
1-pomeron exchange. We further estimate the time it takes for this entropy to be deposited in section 4. In section 5 and 6 we suggest that this stringy entropy is at work in $pp$, $pA$ and $AA$ collisions at collider energies and show that it can account for a key jump in the total charged multiplicities versus the number of participants as reported by the PHOBOS collaboration. The obtained charged multiplicities are compared to the $pp$ and $AuAu$ data.

\vskip0.2cm
{\bf 2.\,\,Stringy Instanton.}
At large rapidity $\chi$ the pomeron exchange can be viewed as a closed string exchange between twisted dipoles~\cite{Basar:2012jb}. At large impact parameter ${\bf b}$, the effective string action is the Polyakov action in $D=2+D_\perp$ dimensions. In Euclidean and T-dual form~\cite{Basar:2012jb}

\bear
S&=&{\sigma_T\over 2}\int_0^T d\tau \int_0^1 d\sigma \left((\partial x^0)^2+(\partial y^1)^2+(\partial x^\perp)^2\right) \nonumber\\
&+&{E\over 2}\int_0^T d\tau \left(y^1\partial_\tau x^0-x^0\partial_\tau y^1\right)\bigg|_{\sigma=0,1}\,,\label{1}
\eear
with 
\be
E=F_{01}=\sigma_T \tanh(\chi/2)\,
\label{2}
\ee
a longitudinal electric field along the $y^1$ direction. The twist angle is played by the rapidity $\chi$. Throughout, $D_\perp=3$ and the string tension is $\sigma_T=1/(2\pi\alpha')$~\cite{Stoffers:2012zw}. 

The semiclassical extrema of (\ref{1}) can be labeled by $k>0$. They follow from the saddle points of (\ref{1}) along $T$ and the world-sheet fields. Explicitly, for $x^\perp={\bf b}\,\sigma$

\be
x^0=R(\sigma)\cos(2\pi k \tau/T)\quad,\quad y^1=R(\sigma)\sin(2\pi k \tau/T)\,,
\label{3}
\ee
with $R(\sigma)=({{\bf b}/\chi})\cosh\left(\chi\left(\sigma-1/2\right)\right)$. The saddle point of (\ref{1}) along the $T$ direction is algebraic, giving $T=2\pi k/\chi$. A similar world-sheet instanton for D-brane scattering was discussed in~\cite{Schubert:2011fz}.

We note that with Euclidean signature, (\ref{2}) refers to a 'magnetic field' along the transverse $01$-direction, so that (\ref{3}) describes a 'cyclotron' motion of the string instanton in the $01$-plane with cyclotron frequency $\omega_k=2\pi k/T$. With Minkowski signature, the motion is  hyperbolic with local acceleration

\be
a(\sigma)=\frac 1{R(\sigma)}={\chi\over {\bf b}}{1\over\cosh\left(\chi\left(\sigma-1/2\right)\right)}\,, 
\label{5}
\ee
which has a maximum at the center of the string, $\sigma=1/2$. Due to this acceleration, the string feels a $\sigma$-dependent Unruh temperature
$T_U(\sigma)={a(\sigma)/2\pi}$ that is maximal at the center with  $T_U={\chi /2\pi {\bf b}}\equiv 1/\beta$.

\vskip0.2cm
{\bf 3.\,\,Entropy.}
The stringy instanton solution (\ref{3}-{4}) reduces the on-shell action (\ref{1}) to

\be
S_k\approx \frac 12 \sigma_k  {\bf b}\beta \ , \label{8}
\ee
with the k-string tension $\sigma_k=k\sigma_T$ for N-ality $k$. For QCD with 3 colors, only the N-alities $k=1,2$ are allowed.  For QCD at large $N_c$, all N-alities up to the integer value of $N_c/2$ are allowed. Only the N-ality $k=1$ is selected in the process of scattering dipoles in the fundamental representation. In section 5, we argue that $k=2$ is released in {\it dense}  $AA$ collisions.

(\ref{8}) receives quantum contributions. For large $\chi$ and ${\bf b}$, the dominant quantum
correction follows from the transverse diffusion of the tachyonic mode in AdS$_{D_\perp}$. To order $1/\sqrt{\lambda}$ in the t' Hooft coupling~\cite{Basar:2012jb,Stoffers:2012zw}

\be
S_k\approx \frac 12 \sigma_k \beta{\bf b}- \frac {2\pi{\bf b}}{\beta}\left(\frac{D_\perp}{12k}-\frac{(D_\perp-1)^2}{8 \sqrt{\lambda}} \right) \ .
\label{9}
\ee
This Euclidean stringy action amounts to a free energy $F_k=S_k/\beta$. It follows that (\ref{9}) carries an entropy

\be
{\bf S}_k\equiv \beta^2\frac{\partial F_k}{\partial \beta}\approx \chi\left(\frac{D_\perp}{6k}-\frac{(D_\perp-1)^2}{4 \sqrt{\lambda}} \right)\,
\label{11}
\ee
or equivalently

\be
{\bf S}_k\approx 2\left(\alpha_{\bf Pk}-1\right)\chi \ .
\label{12}
\ee
For $k=1$, the pomeron intercept is $(\alpha_{\bf P 1}-1) \approx 0.15$ and the entropy per unit rapidity is about $1/3$. 
We note, 

\be
{\bf S}_k\approx  {\rm ln}N_{wee, k}^2 \  , 
\label{13}
\ee
where $N_{wee, k}$ is the total number of wee-dipoles surrounding each of the incoming dipole pairs involved in the collision, see eqn. (36) in \cite{Stoffers:2012zw}. This is to be contrasted with the fully thermal or incoherent expectation of ${\rm ln}{N}$ and the fully Poissonian or coherent expectation of ${\rm ln}\sqrt{{N}}$ with ${N}$ the mean multiplicity number.

Most of this entropy is the result of the tachyon excitation on the string. Indeed, for large impact parameter ${\bf b}$, the Unruh temperature is smaller than the Hagedorn temperature, 

\be
T_U=\frac{\chi}{2\pi {\bf b}}<T_H=\sqrt{\frac{3\sigma_T}{\pi D_\perp}} \ ,
\label{14}
\ee
which translates to ${\bf b}>\chi/(2\pi T_H)$. As the impact parameter is reduced, the Unruh temperature increases, causing the string excitations to exponentiate, leading to a Hagedorn transition. At the Hagedorn point it may be mapped on the Bekenstein-Hawkins (BH) temperature of a microscopic black hole, \cite{Strominger:1996sh, Susskind:1993ws,Horowitz:1996nw,Khuri:2000eq}. 

\vskip0.2cm
{\bf 4.\,\,Formation Time.}
Over what time is the entropy (\ref{12}-\ref{13}) associated to the dipole-dipole collision released? To answer this question, we note that the emergence of an Unruh temperature on the string world-sheet suggests that semiclassically the metric is locally Rindler. Indeed, the line
element associated to the instanton (\ref{3}) in Minkowski signature is

\be
ds^2\approx -a^2\,R^2d(\tau {\bf b})^2+dR^2+dx^{\perp\,2}
\label{16}
\ee
with Rindler time $t(\tau)=\tau {\bf b}$. The Rindler acceleration $a=\chi/{\bf b}$ implies a Rindler horizon ${\bf R}=1/2a$.

We suggest that the prompt release time $t_R$ can be set to be the time when the diffusing string in transverse AdS$_{D_\perp}$ reaches the effective size of the Rindler horizon ${\bf R}$ by analogy with the time it takes to a string to fall on a black hole
~\cite{Susskind:1993aa,Susskind:1993ws}. Indeed, the string diffusion in rapidity causes the transverse string size to increase as~\cite{Stoffers:2012zw}

\be
<x_\perp^2> =\chi \alpha' \equiv {\bf D}_R\,t(1)\ ,
\label{17}
\ee
with the diffusion constant  in Rindler space ${\bf D}_R=\alpha'/(2{\bf R})$~\cite{Susskind:1993aa,Susskind:1993ws}.
Through the last equality, we reinterpret (\ref{17}) as a diffusion in Rindler space over a typical Rindler time $t(1)={\bf b}$. The release entropy time $t_R$ is then set by the condition ${\bf R}^2={\bf D}_Rt_R$ or $t_R=2{\bf R}^3/\alpha'$. For a QCD string with $\alpha'=1/(2\,{\rm GeV})^{2}=(0.1 {\rm fm})^2$ and a typical impact parameter ${\bf b}\sim 10\sqrt{\alpha'}$, this results in $t_R\sim(25\,{\rm fm})/\chi^3$, which is short.

\begin{figure}[b]
  \begin{center}
  \includegraphics[width=8cm]{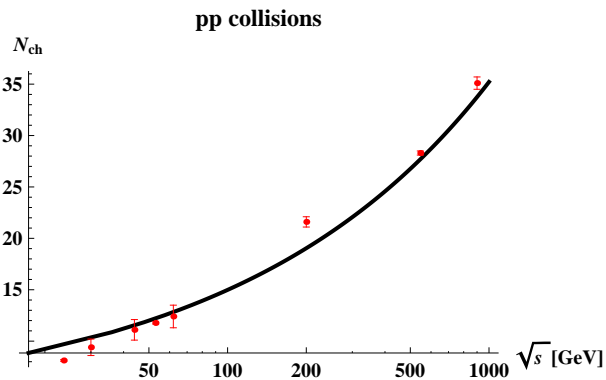}
  \caption{Energy dependence of the charged multiplicity for $pp$ collisions. See text.}
  \label{pp}
  \end{center}
\end{figure}

\begin{figure}[t]
  \begin{center}
  \includegraphics[width=8cm]{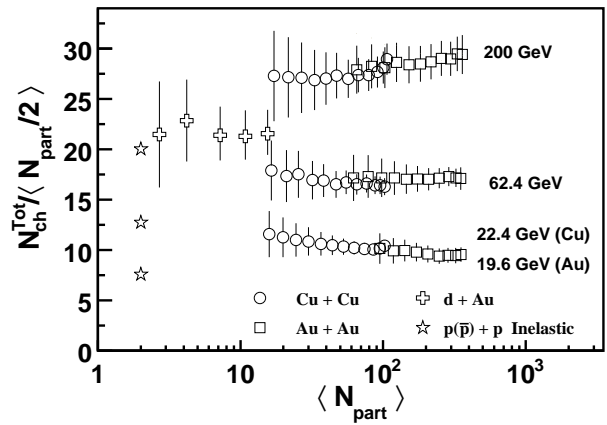}
  \caption{Scaling of the total charged multiplicities with the number of participants~ \cite{Alver:2007aa} . See text.}
  \label{Phobos}
  \end{center}
\end{figure}

\vskip0.2cm
{\bf 5.\,\,pp Multiplicities.}
$pp$ collisions can be viewed as dipole-dipole scattering from each colliding proton \cite{Stoffers:2012ai}. The density of dipoles in the proton is set by the saturation momentum ${\bf Q}_S \equiv \sqrt{2}/z_S$ . In holographic QCD this follows from the transcendental equation~\cite{Stoffers:2012zw}

\be
\frac{z_S}{\sqrt{2}} {\bf Q}_S (\chi , {\bf b}) 
= \frac{g_s^2}{2} \left(2\pi \alpha' \right)^{3/2} \, z_S z_p\,{\bf N}(\chi, z_S,z_p, {\bf b}) =1  \ ,  
\label{transzendental} 
\ee
with the effective string coupling $g_s$ and typical proton virtuality $1/z_p$. The wee-dipole density is set by
\be
{\bf N}=\frac{e^{(\alpha_{\bf Pk}-1) \chi}}{(4 \pi {\bf D_k}\chi )^{3/2}} \left (\frac{1}{z_S z_p} \frac{\xi e^{-\frac{\xi^2}{4{\bf D_k} \chi}}}{\sinh(\xi)} + \frac{z_S}{z_0^2 z_p} \frac{\xi_* e^{-\frac{\xi_*^2}{4{\bf D_k} \chi}}}{\sinh(\xi_*)} \right) \ ,
\ee
with ${\bf D_k}=\alpha'/2k$.  $z_0$ is the confining IR wall. The chordal distances in hyperbolic AdS$_{D_\perp}$ with a wall are 

\begin{eqnarray}
&&{\rm cosh}\xi=1+\frac{{\bf b}^2+(z_S-z_p)^2}{2z_S z_p}\nonumber , \\
&&{\rm cosh}\xi_*=\frac{1}{2}\left( \frac{z_0^2}{z_p z_S}+\frac{z_S z_p}{z_0^2}+\frac{{\bf b}^2 z_S}{z_p z_0^2} \right) \ . 
\end{eqnarray}
The holographic parameters are set by the DIS data analysis in~\cite{Stoffers:2012zw,Stoffers:2012ai}: $\lambda=23$, $D_\perp=3$, $g_s=1.5$, $z_p=1.8  \ {\rm GeV}^{-1}$, $z_0=2 \ {\rm GeV}^{-1}$, $s_0=10^{-2}\ {\rm GeV}^2$. 

If $A_{pp}\approx 1\,{\rm fm}^2$ is the typical proton area, then $A_{pp}{\bf Q}_S^2\approx 12$ is the typical number of dipoles with ${\bf Q}_S^2\approx 1/2\,{\rm GeV}^2$ the typical squared saturation momentum. Thus, for $pp$ collisions the typical entropy release per unit of rapidity is

\be
{\bf S}_{pp}/\chi\approx \left(A_{pp}\,{\bf Q}_S^2\right)\times \left({\bf S}_1/\chi\right)\approx 12\times \frac 13 =4 \ .
\label{19}
\ee

In holography, the scaling of the entropy with the energy follows from the scaling of the saturation momentum, (\ref{transzendental}).  
In the conformal limit and at large $\chi\approx {\rm ln}(s/s_0)$, the entropy asymptotes

\be
{\bf S}_{pp} \approx  \left(\frac s{s_0}\right)^{\left(\sqrt{1+2\sqrt{\lambda}(\alpha_{\bf P}-1)}-1\right)/\sqrt{\lambda}} {\rm ln}(s/s_0) \,
\label{19}
\ee
which is ${\bf S}_{pp}\approx (s/s_0)^{0.228}{\rm ln}(s/s_0)$ using the parameters set by the DIS data.
In Fig. \ref{pp} we show the pp charged multiplicities $N_{ch,pp}={\bf S}_{pp}/7.5$~\cite{Gubser:2008pc} at collider energies~\cite{Basile:1981nt}, with ${\bf b}=1/3\ {\rm fm}$. A recent discussion of the entropy in the context of saturation models was made in~\cite{Kutak:2011rb}.

\vskip0.2cm
{\bf 6.\,\,pA, AA Multiplicities.}
We note that for $pA$ collisions, $A_{pA}\approx A^{1/3}A_{pp}$ so that ${{\bf S}_{pA}}/{{\bf S}_{pp}}\approx A^{1/3}$. In $AA$ collisions, if the collision is mainly between dipoles with N-ality $k=1$, a similar scaling with the nucleus number $A=A^{1/3}\times A^{2/3}$ is expected to take place. Here $A^{1/3}$ Lorentz contracted nucleons  can be distributed in the $A^{2/3}$ transverse nucleus size. However, when the nucleons start to overlap, the $k=2$ N-ality can be exchanged,

\be
\frac{{\bf S}_{AA}}{{\bf S}_{pp}}\approx A\,\left(\sum_1^{[N_c/2]}\frac 1k\right)\, \ .
\label{21}
\ee
In QCD with $N_c=3$, the sum is $3/2$. The contribution of the $k=2$ N-ality is expected to take place when the number of participants is about $10$ so that $10^{1/3}\approx 2$ corresponds to two overlapping nucleons. 

In Fig. \ref{Phobos} we show the total charged multiplicities normalized to the averaged number of participants as a function of the number of participants for a range of collider energies \cite{Alver:2007aa}. For a fixed collider energy, we note the characteristic $3/2$ jump from $pp$ to $AA$ collisions at a number of participant of around 10. 

\begin{figure}[t]
  \begin{center}
  \includegraphics[width=8cm]{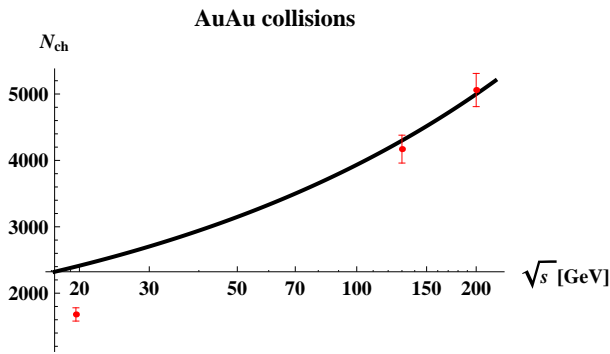}
  \caption{Energy dependence of the charged multiplicity for central $AuAu$ collisions. See text.}
  \label{AuAu}
  \end{center}
\end{figure}

The charged multiplicity follows as $N_{ch,AuAu}=3/2 \langle Au \rangle \ {\bf S}_{pp}/7.5$, with the average participating gold nucleon number $\langle Au \rangle$. Using the same numerical values as for $N_{ch, pp}$ and $\langle Au \rangle =175$ for most central collisions \cite{Back:2002wb}, Fig. \ref{AuAu} shows an agreement of our holographic result with the experimental data  at high energies, where the inelasticities are large. At LHC energies, we expect $N_{ch, pp}\sim 54$, $N_{ch, pPb}\sim 320$, $N_{ch, PbPb}\sim 16800$ at $\sqrt{s}=2.76 \ {\rm TeV}$ and $N_{ch, pp}\sim 82$, $N_{ch, pPb}\sim 470$,$N_{ch, PbPb}\sim 23400$ at $\sqrt{s}=7 \ {\rm TeV}$ using $\langle A_{PbPb}\rangle=191$
\cite{Aamodt:2010cz}.

\vskip0.4cm
\newpage
{\bf 6.\,\,Conclusions.}
We have suggested that the pomeron viewed as an exchange of an instanton on the string world-sheet carries a free energy $F_k/T_U=S_k$ with $S_k$ the instanton action of N-ality $k$ and $T_U$ the Unruh temperature. For large impact parameter ${\bf b}$, the Unruh temperature is low and the entropy is mostly carried by the lowest string excitation, which is tachyonic. This stringy entropy is neither coherent nor thermal.

For smaller impact parameters, the Unruh temperature may  reach the Hagedorn temperature, transmuting the stringy entropy to partonic entropy. The latter is likely commensurate with the Bekenstein-Hawkins entropy, and the onset of a microscopic black hole. 
Macroscopic black holes~ \cite{Shuryak:2005ia,Gubser:2008pc,Lin:2009pn,Wu:2011yd,Kiritsis:2011yn} 
maybe aggregates of these coalescing microscopic black holes as suggested initially in~\cite{Shuryak:2005ia}.

We have argued that typical $pp$, $pA$ and $AA$ collisions at current collider energies may probe this stringy entropy with low Unruh temperature. At large rapidities, the holographic entropy is in agreement with the data for the energy scaling of the charged multiplicities. The 3/2 jump in the charged multiplicities reported by the  current collider experiments with 10 number of participants and higher, is explained by the exchange of N-ality $k=1,2$ strings. We expect similar jumps in the transport parameters, e.g. viscosity and flow.

Although the measured total multiplicities reflect on the final state hadronic produces, entropy conservation garentees that our prompt and initial entropy estimates are lower bounds. The general lore of energy and momentum conservation, say through viscous hydrodynamics evolution, suggests only a moderate increase of the total entropy by about 25\% in going from initial to final states, making our estimates plausible.

\vskip0.2cm
{\bf Acknowledgements.}
We would like to thank Gokce Basar, Dima Kharzeev and Edward Shuryak for discussions.
This work was supported by the U.S. Department of Energy under Contract No.
DE-FG-88ER40388.


\end{document}